\documentstyle[twoside,fleqn,epsf,espcrc2]{article}
\renewcommand{\a}{{\mathrm a}}
\newcommand{\mybar}[1]{\overline{#1}}
\newcommand{\MSbar}{\overline{\mathrm MS}}
\newcommand{\tM}{\tilde{M}}

\title{
       The Charm Quark Mass to Two-Loop Order
			}

\author{
        K.J.~Juge\address{Fermi National Accelerator Laboratory, 
                          P.O.~Box 500, Batavia, Illinois 60510}
			 }

\begin{document}

\begin{abstract}
The truncation of the perturbative series at one loop order for the mass renormalization constants remain a significant systematic uncertainty in the determination of heavy quark masses in lattice QCD. We present here a high beta Monte Carlo calculation of the two loop mass renormalization constant for clover-improved fermions near the charm mass in the Fermilab heavy quark formalism. A preliminary value for the charm quark mass in the $\MSbar$ scheme at two loop order is reported.
\end{abstract}

\maketitle

\section{INTRODUCTION}
Precision calculation of the standard model parameters is one of the goals of lattice QCD. The determination of the charm quark mass, however, remained a difficulty as $am_0\sim1$. 

El-Khadra {\it et.al}\cite{fermilab} have shown that the Wilson action has a stable heavy quark limit. At finite lattice spacing and these heavy quark masses, there are different mass parameters that can be defined with varying lattice artifacts. By appropriately tuning the input parameters and removing the lattice artifacts by continuum limit extrapolation, one can recover the continuum mass definition. The agreement of the different lattice masses has been demonstrated for the charm quark mass by Kronfeld\cite{ask} where the matching to the conventional $\MSbar$ scheme was performed at the one loop level. In this work, we calculate the lattice two loop renormalization constant using large $\beta$ Monte-Carlo techniques which were demonstrated for Wilson quarks in an earlier work\cite{wilson}. These constants are then used to determine the charm quark mass in the $\MSbar$ scheme at two loop order.

\section{CHARM QUARK POLE MASS AND THE $\MSbar$ MASS}

The matching of the lattice to continuum mass is performed using the pole mass as an intermediate step. In determining the pole mass from the lattice at finite lattice spacing, one has several choices for the masses that differ in their lattice artifacts.
The rest mass ($M_1$) and the kinetic mass ($M_2$) were used in Ref.~\cite{ask} to determine the $\MSbar$ mass at one loop order. The rest mass is derived from the spin averaged binding energy of the charmonium 1S states and the other from the quark's kinetic mass. It was shown in that paper that the two quantities defined as such do indeed agree in the continuum limit. In this work, we choose to work with the former quantity, $M_1$, mainly for its simplicity. 

The two loop relation between the pole mass and the $\MSbar$ mass has been calculated by Gray {\it et. al}\cite{broadhurst}. The relationship in the quenched approximation is given by,
\begin{eqnarray}
M&=&\bar{m}(M)\left\{1+\frac{4}{3}\frac{\bar{\alpha}(M)}{\pi}\right.\nonumber\\
&&\,\,\,\,\,\,\,\,\,\,\,\,\,\,\,\,\,\,\,\,\,\,\left.+16.11\left(\frac{\bar{\alpha}(M)}{\pi}\right)^2+{\mathcal O}(\bar{\alpha}^3)\right\}\nonumber
\end{eqnarray}
where $M$ is the pole mass. The corresponding expression for the pole mass determined from the spin averaged binding energy is\cite{ask},
\begin{eqnarray}
M&=&\frac{1}{2}\left\{\mybar{M}^{expt}_{c\bar{c}}-\a^{-1}(\mybar{\a M}_{1c\bar{c}})^{MC}+2\a^{-1}\tM^{[0]}_1\right\}\nonumber\\
&+&\alpha_V(q^\star)(\a^{-1}\tM^{[1]}_1)+\alpha_V(q^\star)^2(\a^{-1}\tM^{[2]}_1)\nonumber
\end{eqnarray}
Note that the definition of the perturbative coefficients ($\tM^{[i]}_1$) is slightly different than those given in Ref.\cite{bart}. 

The charm quark mass in the $\MSbar$ scheme is determined at two loop order from these two expressions.

\section{PERTURBATIVE COEFFICIENTS IN THE LATTICE THEORY}

Mertens {\it et al.}\cite{bart} have tabulated the mass dependent one loop coefficients, $\tM^{[1]}_1$, for a wide range of bare quark masses. We determine the two loop coefficent, $\tM^{[2]}_1$, using large $\beta$ Monte Carlo methods\cite{dimm}. The procedure is essentially the same as described in Ref.~\cite{wilson} except that we use a different gauge and periodic twisted boundary conditions. We first fix to the axial gauge instead of starting from the Coulomb gauge for these clover-improved quarks. The periodic twisted b.c. gives much smaller finite volume dependence than the anti-periodic b.c. used for the Wilson quarks. 

\subsection{Simulation}

We use three values of the hopping parameter which were tuned to the spin-averaged $c\bar{c}$ kinetic mass. The three values correspond to tuning performed at $\beta=5.7$, 5.9 and 6.1. Mean link tadpole-improved tree level value for the clover coefficient was used in the tuning. Since we do not know the nonperturbative value of the critical mass at large $\beta$, the tadpole-improved bare quark mass ($\tilde{m}_0$) is held fixed in the perturbative expansion. This gives rise to a difference in the treatment of the power divergences between this work and Ref.~\cite{bart} where this is treated nonperturbatively. However, the knowledge of the two loop constant for the critical mass will allow to switch between the two schemes.

The couplings and the lattice volumes used are similar to the ones used for the Wilson quark study. This includes 7 values of $\beta\ge 9$ and 7 volumes, from $4^3\times 16$ to $16^4$. The statistics is comparable to the anti-periodic lattices. For $\beta=60$, we have an extra lattice, $24^4$, with lower statistics to provide a check for the infinite volume extrapolation. 

\subsection{Fitting Procedure}
 
The quark propagators are fit to a hyperbolic cosine from timeslice 3 to extract the rest mass, $M_1$. These were all good fits with $\chi^2$'s less than the degrees of freedom. The fits from timeslice 4 and 5 gave results which were consistent but with larger errors.

At each weak coupling, we extrapolate the rest masses to the infinite volume limit where we assume the following form for the finite volume corrections,
$$
V_{int}(L) = v_1\frac{1}{L}+v_2\frac{\ln(L)}{L}
$$
where $v_1$ and $v_2$ are varied independently. 

We then fit the rest masses from each weak coupling to a power series (fourth order) in $\alpha_V(q^\star)$. The prescription for the determination of $q^\star$ given in Ref.\cite{latpert} results in a very large $q^\star$ especially for $\kappa=0.126$ since the one-loop coefficient is rather small. However, we note that the two loop coefficients extracted at two different reasonable $q^\star$s scales within the errors. We therefore use the values of the BLM-LM combined $q^\star$s from Ref.~\cite{ask}. 

We summarize the result of the fits without fixing any of the coefficents in Table~\ref{oneloop}. We are able to reproduce the one loop coefficients within the statistical errors. 

\begin{table}
\begin{center}
\begin{tabular}{|c|c|c|c|c|}\hline
$\kappa$ & $\tilde{M}^{[0]}_1$ & $\tilde{M}^{[1]}_1$ from \cite{ask} &  $\tilde{M}^{[1]}_1$ large $\beta$\\\hline
0.119  & 0.6318 & 0.4502 & 0.43(2)\\
0.1227 & 0.5102 & 0.3195 & 0.30(2)\\
0.126  & 0.4061 & 0.1818 & 0.20(2)\\\hline
\end{tabular}
\end{center}
\caption[]{Fit results without any constraints.}
\label{oneloop}
\end{table}

\subsection{Constrained Curve Fitting}
The higher order terms in the unconstrained fit produces coefficients that are much larger than the one-loop term, but also has large errors such that they are consistent with zero. To determine the two loop coefficients, we use a constrained fitting method using Gaussian priors\cite{bayes} to reduce the errors. The one-loop term is constrained to roughly 1\% of the known value. The higher order terms have been constrained with a central prior of 0 with a width of 100. The fits are stable against the choice of the widths of the priors, but we note that the (regular) $\chi^2/d.o.f.$ for $\kappa=0.126$ is somewhat large ($>3$). The results are in Table~\ref{constrain}.

\begin{table}
\begin{center}
\begin{tabular}{|c|c|c|c|}\hline
$\kappa$ & $\tilde{M}^{[0]}_1$ & $\tilde{M}^{[1]}_1$ &  $\tilde{M}^{[2]}_1$ \\\hline
0.119  & 0.6318 & 0.452(7) & 1.3(6)\\
0.1227 & 0.5102 & 0.320(6) & 1.4(5)\\
0.126  & 0.4061 & 0.185(3) & 2.1(4)\\\hline
\end{tabular}
\end{center}
\caption[]{Two-loop results from the constrained fits. Only statistical errors are quoted.}
\label{constrain}
\end{table}

\begin{figure}
\begin{center}
\epsfxsize=2.5in \epsfbox{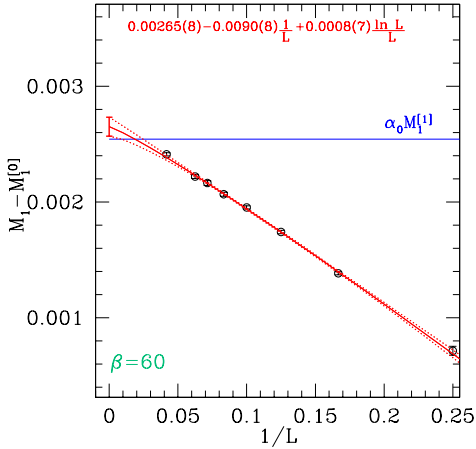}
\end{center}
\vspace{-12ex}
\caption[]{The infinite volume extrapolation of $M_1-M_1^{[0]}$ at $\beta=60$. Note that the point $L=24$ was not used in the fit.}
\label{b60}
\end{figure}

\begin{figure}
\begin{center}
\epsfxsize=2.5in \epsfbox{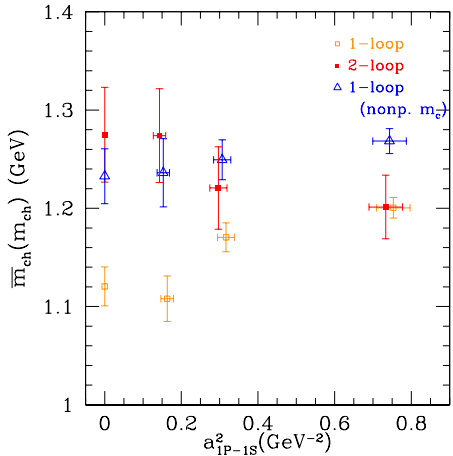}
\end{center}
\vspace{-12ex}
\caption[]{The 1-loop and 2-loop $\MSbar$ masses at finite lattice spacings. The continuum limit is taken with $a^2$ errors.}
\label{charm}
\end{figure}

\section{RESULTS/DISCUSSION}
We use the two loop results obtained above from large $\beta$ Monte-Carlo  to calculate the charm quark mass in the $\MSbar$ scheme at two loop order with given lattice cutoffs. The masses are extrapolated to the continuum limit using $a^2$ errors. The results for the one-loop, both in this work and Ref.~\cite{ask}, and the two loop $\MSbar$ mass are summarized in Fig.~\ref{charm}. Note that the difference in treatment of $m_c$ is significant at the one loop level. Our preliminary result for the two-loop charm quark mass in the $\MSbar$ scheme is $1.27(5)$ where only the statistical errors are quoted. The fitted two loop values are fairly stable against the choice of $q^\star$, truncation of the series and the widths of the priors. The exception is the lightest bare quark mass where the one loop coefficient is so far not reproducable without a sizeable $\chi^2$. The infinite volume extrapolation is still under study as well as the conversion to the nonperturbative treatment of the critical mass.                  
\section*{ACKNOWLEDGEMENTS}
This work was done in collaboration with A.~Kronfeld, P.~Lepage,  P.~Mackenzie, J.~Simone and H.~Trottier. 
Fermilab is operated by University Research Association, Inc. for the U.~S.~Department of Energy.



\begin{thebibliography}{9}
\bibitem{fermilab}
   A.X.~El-Khadra, A.S.~Kronfeld and \\P.B.Mackenzie, Phys.\ Rev.\ {\bf D 55}, 3933 (1997).
\bibitem{ask}
   A.S.~Kronfeld, Nucl.\ Phys. B (Proc.\ Suppl.) {\bf 42}, 403 (1997) \bibitem{wilson}
   K.J.~Juge, Nucl.\ Phys. B (Proc.\ Suppl.) {\bf 42}, 403 (2000) 
\bibitem{broadhurst}
   N.~Gray {\it et al.} Z.~Phys.~C {\bf 48}, 673 (1990) 
\bibitem{bart}
   B.P.G.Mertens, A.S.Kronfeld and A.X.El-Khadra, Phys.\ Rev.\ {\bf D 58}, 034505 (1998).
\bibitem{dimm}
   W.Dimm, G.P.Lepage and P.B.Mackenzie, Nucl.\ Phys. B (Proc.\ Suppl.) {\bf 42}, 403 (1995).
\bibitem{latpert}
   G.P.~Lepage and P.B.~Mackenzie, Phys.\ Rev.\ {\bf D 48}, 2250 (1993).
\bibitem{bayes}
	G.P.~Lepage, these proceedings.
\end{thebibliography}
\end{document}